\documentclass[a4paper]{article}
\usepackage{amsmath,amssymb}
\title{Public key cryptosystems based on Iterated Functions Systems}
\author{Jacques Peyri\`{e}re$^{~a,b}$, ~~ Fengxia Liu$^{*c,d}$ ~~ Zhiyong Zheng $^{a}$~~Zixian Gong $^{a}$  \\[10pt]
\small {$^{a }$ Engineering Research Center of Ministry of Education for Financial Computing }\\
\small {and Digital Engineering, Renmin University of China, }\\
\small {  Beijing, 100872,  P. R. China}\\[5pt]
\small { $^{b }$Institut de 
Mathématiques d’Orsay, CNRS, Université Paris-Saclay, 91405 Orsay, France}\\[5pt]
\small { $^{c }$Institute of Artificial Intelligence, Beihang University, Beijing, 100191,  P. R. China}\\[5pt]
\small { $^{d }$    Beijing Advanced Innovation Center for Future Blockchain and Privacy Computing, Beijing,  P. R. China}\\
\\[5pt]}

\footnotetext{*~Corresponding author (Fengxia Liu).\\
E-mail addresses:  shunliliu@buaa.edu.cn }

\date{August 27, 2023 (8th draft)}
\newcommand{\Rem}{{\mathrm{rem}_p}}
\newcommand{\receive}{{\ $\mathbf<$\hspace{-4pt}\textbf{---}\ }}
\newtheorem{definition}{Definition}
\newtheorem{algorithm}{Algorithm}

\begin{document}

\maketitle

\begin{abstract}
Let $f=(f_0,f_1,\dots, f_{\nu-1})$ be a collection of one-to-one functions from some space~$X$ into itself such that the sets $f_j(X)$ are disjoint. If $w=w_1w_2\cdots w_k$ is a word on the alphabet $\{0,1,\dots,\nu-1\}$, let $\Phi_{f,w} = f_{w_1}\circ f_{w_2}\circ\cdots\circ f_{w_k}$. Given a function~$F$ of which we know that it can be written as $\Phi_{f,w}$, it is easy to  recover~$w$. We give some examples of this situation where everything can be scrambled up by using some private key to get a new system $g=(g_1,g_2,\dots,g_{\nu-1})$ on another set~$Y$ in such a way that the images of the $g_j$ are no longer disjoint. We define a cryptosystem whose public key is~$g$. The message to be encrypted is a word~$w$ and the associated cryptogram is $\Phi_{g,w}$. The private key allows to recover $\Phi_{f,w}$ from $\Phi_{g,w}$.
\end{abstract}


\section{Introduction}

Let $f = (f_0,\,f_1,\dots,\,f_{\nu-1})$ be a collection of mappings from a set~$X$ into itself. If $w = w_1w_2\cdots w_n$ is a word over the alphabet $\{0,1,\cdots,\nu-1\}$ we set
\begin{equation}\label{ifs}
	\Phi_{f,w} = f_{w_1}\circ f_{w_2}\circ\cdots\circ f_{w_n}.
\end{equation}
Such a system is called an IFS (Iterated Functions System).
\medskip

The IFS have been introduced by Hutchinson~\cite{hutchinson} to rationalize several preexisting constructions of fractal sets and measures having some self-similarity properties. This formalism has been adopted by the fractalists' community and gave rise to an abundant literature.  Subsequently this formalism has been further developed by Mauldin and Williams~\cite{williams} by the so called ``Graph directed constructions''.

\begin{definition}\label{separation}
The system~$f$ is said to have the separation property relatively to ${\mathcal S} = (S;S_0,S_1,\dots,S_{\nu-1})$ if
\newpage
\begin{itemize}
	\item [--] $S\subset X$, $S\ne \emptyset$,  and $S_j\subset S$ for all~$j$,
	\item [--] for all $j$, $f_j$ defines a bijection  from~$S$ onto~$S_j$,
	\item [--] $S_i\cap S_j = \emptyset$ for $0\le i< j< \nu$.
 \end{itemize}
\end{definition}
\medskip

Several separation conditions have been defined for IFS which aimed at computing or estimating the Hausdorff dimension of the limit set of the said IFS. This is not our present goal and the separation condition we consider does not coincide with the previous ones, although the closest one is the so-called ``strong separation condition''. Indeed the system $(x/2,\ 1-x/2)$ satisfies our condition on the interval~$[0,1)$ and also on the interval~$(0,1]$, while it satisfies on~$[0,1]$ the ``open set separation condition'' which is important un fractal geometry. But for our purpose, this is our definition which is meaningful. And in the sequel, when we say that a system satisfies (or has, or fulfills) a separation condition this means that there exists an~${\mathcal S} = (S;S_0,S_1,\dots,S_\nu)$ as in definition~\ref{separation}.

When we have an IFS we can consider the following two problems which are closely related and which are easily solved when this IFS satisfies a separation properties.

\noindent
\textbf{Problem 1}. ``Given an integer~$n$ and a point~$y\in S$, can we find $w\in \{0,1,\dots,\nu-1\}^n$ such that there exists $x\in S$ such that $y=\Phi_{f,w}(x)$?''.

Indeed, here is the solution.
If $y \notin \bigcup f_j(S)$ there is no solution. Otherwise, if $y\in f_j(S)$, then, if a solution exists, we must have $w_1=j$. Then we are confronted with the same problem: to find~$w'$ of length~$n-1$ such that $\Phi_{f,w'}(x) = f_{j}^{-1}(y)$, where $f_{j}^{-1}$ is the reciprocal of the restriction of~$f_{j}$ to~$S$. Here is a description of this algorithm.\bigskip

\begin{algorithm} \label{algo1}{\ }\\	
	{\noindent
		w\receive\  empty word\\
		while $|w|<n$ repeat\{\\
		\hspace*{1em} if $y\notin S_1\cup S_2\cup\dots\cup S_{\nu-1}$ then return ``failed''\\
		\hspace*{1em} if $y\in S_j$ then \{\\
		\hspace*{2em} $w$\receive $wj$\\
		\hspace*{2em} $y$\receive $f_j^{-1}(y)$\\
		\hspace*{1em}\}\\
		\}\\
		return $w$
	}
\end{algorithm}

\noindent
\textbf{Problem 2}. ``Given two points~$x\in S$ and~$y\in S$ of which we know that $y=\Phi_{f,w}(x)$ for some word~$w$ on the alphabet $\{0,1,2,\dots,\nu-1\}$, find this word~$w$''. Here is an algorithm which does the job:
\begin{algorithm} \label{algo2}	
	{\noindent
		w\receive\  empty word\\
		z\receive\ y\\
		while $z\ne x$ repeat\{\\
		\hspace*{1em} if $z\notin S_1\cup S_2\cup\dots\cup S_{\nu-1}$ then return ``failed''\\
		\hspace*{1em} if $z\in S_j$ then \{\\
		\hspace*{2em} $w$\receive $wj$\\
		\hspace*{2em} $z$\receive $f_j^{-1}(z)$\\
		\hspace*{1em}\}\\
		\}\\
		return $w$
	}
\end{algorithm}
The main difference between these problems is that the length of the unknown word~$w$ is not specified in the second one.
\bigskip

To get a cryptosystem from an IFS~$f$ with the separation condition, the strategy is to construct by some scrambling method a set of functions $g=(g_0,\,g_1,\dots,\,g_{\nu-1})$ related to~$f$ but not obviously fulfilling a separation condition. This strategy will be developed in the next sections.

The graph directed constructions give rise to more complex systems of iteration of functions. In the same way as the ones described below they can give rise to cryptosystems, but we will not elaborate on this subject. 
	
\section{First protocol: Affine cryptosystems}

\subsection{Preliminaries and notation}

Let~$p$ be a prime number.
Let~$\varphi$ be the canonical projection of~$\mathbb Z$ onto the quotient ${\mathbb F}_p := {\mathbb Z}/p{\mathbb Z}$.  For the sake of precision let us have a notation for the usual section of~$\varphi$: 
for $\alpha\in {\mathbb F}_p$, let $\Rem(\alpha)$ stand for the integer $m\in \{0,1,2,\dots,p-1\}$ such that $\varphi(m) = \alpha$.\bigskip

Consider the following sub-ring ${\mathcal R}_p$ of~${\mathbb Q}$:
\begin{equation}\label{ring}
	{\mathcal R}_p = \left\{r\in {\mathbb Q}\ :\ \exists m\in {\mathbb Z}, \exists n\ge 1 \text{ such that } (n,p) = 1 \text{ and } r=m/n \right\}.
\end{equation}
For $m\in {\mathbb Z}$ and~$n>0$ such that $(n,p)=1$, set $\widetilde{\psi}(m,n) = \varphi(m)\varphi(n)^{-1}$. If we have another couple $(m',n')$ such that $m'/n'=m/n$, we have
$$ \widetilde{\psi}(m,n) = \varphi(mn')\varphi(n')^{-1}\varphi(n)^{-1} = \varphi(m'n)\varphi(n')^{-1}\varphi(n)^{-1} = \widetilde{\psi}(m',n').
$$
This means that if $r=m/n\in {\mathcal R}_p$, it is legitimate to set $\psi(r) = \widetilde{\psi}(m,n)$. One can check that $\psi$ is a ring homomorphism from ${\mathcal R}_p$ to ${\mathbb F}_p$ which extends~$\varphi$.

As usual, $\psi$ is extended as a ring homomorphism from ${\mathcal R}_p[x]$ to ${\mathbb F}_p[x]$.
When~$A=\sum_{j=0}^k a_j x^j$ and~$B$ are polynomials on some ring the polynomial $A(B)$ or $A\circ B$ is defined to be the polynomial $\sum_{j=0}^{k}a_j B^j$. It is clear that for polynomials~$A$ and~$B$ in~${\mathcal R}_p[x]$ that we have $\psi(A\circ B) = \psi(A)\circ\psi(B)$.
In particular if $A\in {\mathcal R}_p[x]$ and $b\in {\mathcal R}_p$ we have 
$\psi\bigl(A(b)\bigr) = \psi(A)\bigl(\psi(b)\bigr)$.
\medskip

Using $\varphi,\,\psi$, and~$\Rem$ may seem pedantic. Nevertheless this has the merit to be precise. This is why we use this notation in the next example. But afterwards we adopt a more familiar convention. Depending on the context   \mbox{`\hspace{-5pt}$\mod p$`} will have several meanings: if~$q\in {\mathcal R}_p[x]$, then \quad $q \mod p$\quad could stand for $\varphi(q)\in {\mathbb  F}_p[x]$, $\psi(q)\in {\mathbb F}_p[x]$,  or for the corresponding polynomial in ${\mathbb Z}[x]$ whose coefficients lie between~$0$ and~$p-1$.

\subsection{An example}

Let us begin with an example. Take two positive integers $\alpha$ and~$\beta$. Let $f_0(x) = \frac{x+2\alpha}{3}$ and $f_1(x) = \frac{x+2(\alpha+\beta)}{3}$.	Observe that this system $f=(f_0,f_1)$ fulfills the separation condition in the interval~$S := [\alpha,\alpha+\beta]$, i.e.,
$${\mathcal S} = \left([\alpha,\alpha+\beta];\left[\alpha,\alpha+\frac{\beta}{3}\right],\left[\alpha+\frac{2\beta}{3},\alpha+\beta\right]\right).$$
\bigskip

The plain text message to be encrypted will be a word $w$ of length~$n$ on the alphabet~$\{0,1\}$. Observe that, for such a word~$w$, we have $\Phi_{f,w}(\alpha) = \alpha+ m3^{-n}\in [\alpha,\alpha+\beta]$, where~$m$ is an integer. It results that $0\le m\le 3^n\beta$.

\subsubsection*{Key generation}
First, choose a prime number~$p> 3^n\beta$. Then take two different coprime numbers~$a$ and~$b$ such that $2\le a,\, b< p$. Set $u(x) = ax+b$. Then the reciprocal function of~$u$ is $u^{-1}(x)= (x-b)/a$. 
We have $u^{-1}\circ f_0\circ u \in {\mathcal R}_p[x]$ (see~\eqref{ring}), and the same for~$f_1$. So we define the IFS $g=(g_0,g_1)$ on ${\mathbb F}_p$ as follows:
$$g_\varepsilon = \psi\left(u^{-1}\circ f_\varepsilon\circ u \right) \text{\quad for }  \varepsilon\in \{0,1\}.$$
In theses conditions we have
\begin{equation}
	\Phi_{g,w} = \psi\left(u^{-1}\circ \Phi_{f,w}\circ u\right). \label{main}
\end{equation}
%
%
Define $\gamma = \phi\left(	u^{-1}(\alpha)\right)$. Let the public and private keys be
\begin{equation}
 pk = \{n,p,\,\gamma,\,g\} \text{\quad and\quad} sk = \{n,\,p,\,u,\, f, {\mathcal S}\}.
\end{equation}

\subsubsection*{Encryption}
If Bob wishes to send Alice the message~$w$ of length $n$, he computes 
$$c = \Phi_{g,w}(\gamma)$$
and sends Alice the cryptogram $c$.

\subsubsection*{Decryption}
To decipher this cryptogram Alice first compute $u(c)$. Due to~\eqref{main} we have $\psi\bigl(\Phi_{f,w}(\alpha)\bigr) = u(c)$. Therefore we have $\psi\Bigl(3^n\bigl(\Phi_{f,w}(\alpha)-\alpha\bigr)\Bigr) = 3^n\left(u(c)-\varphi(\alpha)\right)$ (this is an equality between elements of ${\mathbb F}_p$). But, because $3^n\bigl(\Phi_{f,w}(\alpha)-\alpha\bigr)$ is an integer less than or equal to~$3^n\beta<p$, we have $3^n\Phi_{f,w}(\alpha) = \alpha+\Rem\bigl(3^nu(c)-3^n\varphi(\alpha)\bigr)$. Finally we obtain
$$\Phi_{f,w}(\alpha) = \alpha+\frac{\Rem\bigl(3^nu(c)-3^n\varphi(\alpha)\bigr)}{3^n}, $$
which allows to use the algorithm~\ref{algo1} to get~$w$.

\subsubsection*{Remarks} 

First we notice that the new system $g$ does not exhibit an obvious separation property. Indeed we illustrate this fact by considering the following simple example. $n=8$, $p=19687$, $f_0(x)=x/3$, $f_1(x)=(2+x)/3$. $a=15296$, $b=8026$. Then $g_0(x) = 13125x + 8750 \mod p$ and  $g_1(x) = = 13125x + 10515 \mod p$. Then there are four minimal invariant sets $E_{0,0},E_{0,1},E_{0,2},E_{0,3}$ under $g_0$. The set $E_{0,0}$ has one element only. The other ones have 6562 elements. The same siyuation occurs for $g_1$: we have the invariant sets $E_{1,0},E_{1,1},E_{1,2},E_{1,3}$ with the same cardinality as previously. All the intersections $E_{0,i}\cap E_{1,j}$ are non empty for $1\le i,\, j<3$. This means that there is no non trivial set invariant under~$g_0$ and~$g_1$. As a consequence the system~$g=(g_0,g_1)$ does not satisfy any separation condition. So to break this cryptosystem one is left with the brute force attack. As the key length is much larger that the length of the messages, the attack consists in feeding the enccryption algorithem with all the possible messages until obtaining the cryptogram. This requires $O(2^n)$ operations ($n$ being the length of the message).

As we just realized, an eavesdropper cannot easily decode the cryptogram, but if he has the ability to usurpate  Bob's identity he can tamper with the cryptogram~$c$, for instance by replacing~$c$ by~$g_0(c)$ and send the modified cryptogram to Alice. When decoding there are two possibilities.
\begin{enumerate}
	\item The decoding fails, and Alice knows that the cryptogram has been tampered with.
	\item The decoding works for n steps. Then the cryptogram is wrongly deciphered and Alice is not aware of it. 
\end{enumerate} 

It is not difficult to obviate this problem. This time the message to be transmitted is a binary word~$w$ of length less than or equal to~$n$. We choose a prime number~$p>3^{n+\lceil\log_2 n\rceil}\beta$. The keys are constructed as previously. To make the cryptogram associated with the message~$w$ Bob appends to~$w$ a dyadic word of length $\lceil\log_2 n\rceil$ containing the dyadic expansion of~$|w|$ (where $|w|$ stands for the length of~$w$) padded on the left by zeros, if necessary. Let the resulting word  be~$w'$. Then the cryptogram is $c = \Phi_{g,w'}(\gamma)$.

As previously, the knowledge of $u(c)$ gives the value of $\Phi_{f,w'}(\alpha)$. Alice first recovers~$w'$ by applying the algorithm~\ref{algo2} to $\Phi_{f,w'}(\alpha)$. Then she obtains $w$ by removing the added rightmost bits. Moreover since she knows the length of~$w$ she is able to detect an adulteration.

\subsection{A general framework}

We deal with an IFS $f=(f_j)_{0\le j< \nu}$ consisting in~$\nu$ affine maps from~${\mathbb Q}$ to~${\mathbb Q}$. We suppose that this IFS satisfies the separation condition on an interval~$[\alpha,\alpha+\beta)\subset [0,+\infty)$. Let~$\varpi$ be the $\mathrm{lcm}$ of the denominators  of the coefficients of the $f_j$.
Then if $w\in \{0,1,\dots,\nu-1\}^n$ we have $\Phi_{f,w}(\alpha) = \alpha+k/\varpi^n$, where $0\le k\le \varpi^n\beta$.\medskip

Choose a prime number $p> \beta\varpi^{n+\lceil \log_\nu n\rceil}$ and two coprime numbers~$a$ and~$b$ less than~$p$. Set $u(x)=ax+b$, $v(x)=(x-a)/b$, $\gamma = v(\alpha) \mod p$, and $g = \left(g_j\right)_{0\le j<\nu}$, where $g_j = v\circ f_j\circ u \mod p$ (for $0\le j<\nu$).
 Let the keys to be
 $$pk = (n,p,\gamma,g)\text{\quad and \quad} sk = (n,p,u,f,{\mathcal S}).$$
 
 The cryptogram~$c$ attached to the message~$w\in\{0,1,\dots,\nu-1\}^m$ with $m\le n$ is constructed as follows: let $w'$ the word obtained by appending to~$w$ the base-$\nu$ expansion of~$m$, padded on the left by zeros, if necessary. Then $c = \Phi_{g,w'}(\gamma) \mod p$.
 
 The decryption runs as follows: Alice obtains $\Phi_{f,w'}(\alpha)$ as
 $$\alpha + \frac{\varpi^n\bigl(u(c)-\alpha\bigr) \mod p}{\varpi^n}$$ 
 and uses the algorithm~2 to get~$w'$. Then she obtains~$w$ by removing from~$w'$ the digits added to the right.

\subsubsection{Remark}

Of course, one can use systems of affine maps in higher dimension.

\section{Second protocol: Projective cryptosystems}
 
\subsection{Preliminaries}
 
\subsubsection*{Projective spaces and homographies}
 
The  projective space of dimension~$d$ on some field~$K$, denoted by ${\mathbb P}_d(K)$, is the set of one dimensional vector subspaces of~$K^{d+1}$. Any $x=(x_1,x_2,\dots,x_{d+1})\in K^{d+1}\setminus \{0\}$ determines a 1-D vector space, i.e., a point of the projective space; call it $P_x$. Obviously, if $\lambda\in K^*$, $\lambda x = (\lambda x_1,\lambda x_2,\dots,\lambda x_{d+1})$, then $P_{\lambda x} = P_x$. 
 
If $P =P_x$ we say that~$x$ is a set of homogeneous coordinates for~$P$. The point~$P_x$ has many sets of homogeneous coordinates: they are all the $\lambda x$ where~$\lambda$ is any non-zero element of~$K$. The affine space $K^d$ is embedded in ${\mathbb P}_d(K)$: this is the set of points of which the last term of any of their system of homogeneous coordinates is non-zero; in other terms, a point $(x_1,x_2,\dots,x_d)\in K^d$ is identified with the point of homogeneous coordinates $(x_1,x_2,\dots,x_d,1)$. Of course there are many embeddinds of the affine space $K^d$ in ${\mathbb P}_d(K)$.

 A non-zero endomorphism~$u$ of $K^{d+1}$ induces a mapping, called an \emph{homography} of ${\mathbb P}_d(K)$, indeed $u$ transforms as set projective coordinates into a set of projective coordinates of the image. Obviously, $u$ and $\lambda u$ (for $\lambda \in K^*$) induce the same  homography.
 
 \subsubsection*{Projective IFS}
 
 We consider an IFS $f = (f_0,f_1,\dots,f_{\nu - 1})$ whose elements are invertible homographies on ${\mathbb P}_d(K)$. This means that we have~$\nu$ invertible $(d+1)\times(d+1)$-matrices $A_0,A_1,\dots,A_{\nu-1}$ with coefficients in~$K$.
 If $w=w_1w_2\cdots w_n$ is a word ~$n$ on the alphabet $\{0.1.2.\dots,\nu-1\}$ we set
 $$
 \Phi_{A,w} = A_{w_1}A_{w_2}\cdots A_{w_n},
 $$
 where $A$ stands for $(A_0,A_1,\dots,A_{\nu-1})$. Obviously $\Phi_{f,w}$ is the homography associated with the matrix $\Phi_{A,w}$.\medskip
 
  In case when this IFS fulfills a separation condition the following problem is easy to solve.\medskip

 \noindent\textbf{Problem3:} Given a matrix~$M$  of which we know that there exist an integer~$k$ and a word $w\in \{	0,1,2,\dots,\nu-1\}^k$ such that $M = \Phi_{A,w}$, recover the word~$w$.\medskip
 
 \noindent\textbf{Solution:} Choose a point~$P$ in~$S$ (we stick to the notation of the introduction); more precisely we choose a system of homogeneous coordinates of~$P$ that define a column~$X$. We certainly have $f_{w_j}\circ f_{w_{j+1}}\circ \dots\circ f_{w_k}(P)\in S$ for all~$j\le k$. So there exists~$j$ such that $MX\in S_j$ (with a slight abuse of notation). This means that~$w_1=j$. And now we have to solve the same problem with $A_j^{-1}M$. We then repeat this step until reaching the unit matrix.\bigskip

 \begin{algorithm} \label{algo3}{\ }\\	
 	{\noindent
 		choose $P\in S$\\
 		w\receive\  empty word\\
 		while $M\ne I$ repeat\{\\
 		\hspace*{1em} if $MP \in S_j$ then \{\\
 		\hspace*{2em} $w$\receive $wj$\\
 		\hspace*{2em} $M$\receive $A_j^{-1}M$\\
 		\hspace*{2em}\}\\
 		\hspace*{1em}\}\\
 		return $w$
 	}
 \end{algorithm}\medskip

\textsl{\bfseries Examples}\\[.3em]

The system consisting in the homographies defined by these two maatrices $\begin{pmatrix} 1&0\\1&1\end{pmatrix}$ and $\begin{pmatrix} 1&1\\0&2
\end{pmatrix}$ fulfills the separation condition on the interval~$[0,1)$.\\
The system defined by the matrices
$\begin{pmatrix} 1&0&0\\0&1&0\\1&1&1 \end{pmatrix}$, $\begin{pmatrix}1&0&1\\0&1&1\\0&0&2\end{pmatrix}$, 
$\begin{pmatrix} 1&0&1\\0&1&0\\0&0&2\end{pmatrix}$, and $\begin{pmatrix} 1&0&0\\0&1&1\\0&0&2\end{pmatrix}$ fulfills the separation condition on the square~$[0,1)\times[0,1)$.
\medskip

\subsection{Cryptosystems}

 Now we consider  a set $A = (A_0,A_1,\dots,A_{\nu-1})$ of invertible $(d+1)\times(d+1)$-matrices with integer entries which defines a homographic IFS fulfilling the separation condition on some~$\mathcal S$. We may suppose that the coefficients of eachs $A_j$ have no common factor.
 \medskip
 
 \subsubsection*{Key generation}
 
Choose $n$ (the maximum length of the messages to be encrypted) a prime number~$p$ larger than twice
the absolute value of any coefficient of any $\Phi_{A,w}$ with $|w|\le n$ (see below how to make such a choice),
and a square matrix~$U$ of dimension~$d+1$, with integer coefficients, and whose determinant is not a multiple of~$p$.
Let $B_i = U^{-1}A_jU \mod p$ and $B=(B_0,B_1,\dots,B_{\nu-1})$. Let the public and private keys be:
$$ pk = \left(n,p,B\right) \text{\quad and\quad} sk = \left(n,p,A,U,{\mathcal S}\right).$$

\subsubsection*{Encryption}

The message to be transmitted is a word~$w$ of length $k\le n$ on the alphabet~$\{0,1,2,\dots,\nu -1\}$. The corresponding cryptogram is the matrix
\begin{equation*}
	C = B_{w_1}B_{w_2}\cdots B_{w_k} \mod p.
\end{equation*}

\subsubsection*{Decryption}

We have $A_{w_1}A_{w_2}\cdots A_{w_k} = UCU^{-1} \mod p$.\\ But all the coefficients of $A_{w_1}A_{w_2}\cdots A_{w_k}$ have an absolute value less than $p/2$. So we have 
$$A_{w_1}A_{w_2}\cdots A_{w_k} = UCU^{-1} \text{\ \ mods\ \  }p,$$
where $a\text{\ \ mods\ \  }p$ stands for the number~$b$ such that $|b|<p/2$ and $b = a \mod p$.
Then we apply the algorithm~\ref{algo3} to recover~$w$..

\subsubsection*{Remarks}

As previously the brute force attack requires $O(\nu^n)$ operations. Also, as previously  an eaves dropper with the ability of usurpating Bob's identity  can tamper the cryptogram by  pre or post multiplying~$C$ by some of the $B_j$. To thwart this tampering one can put extra information in the middle of the message~$w$, for instance its length. In the next section we show how this protocol allows message authentication and integrity. 

This procol seems to be not well suited, although usable, when all the homographies are affine maps. Indeed in this case the matrices $B_j$ have a common eigendirection, which could weaken the cipher. This is specially the case when they are similitudes on a one-dimensional space..

In section~3.2 we need a bound for the coefficients of a product of matrices. For reader's convenience, we give a possible way to get such a bound. 

We can define two norms on the space of real $(d+1)\times (d+1)$-matrices. If $M=(m_{i,j})_{1\le i,j\le d+1}$ then
$$\|M\|_h = \max_{1\le i\le d+1} \sum_{1\le j\le d+1} |m_{i,j}| \text{\quad and\quad} \|M\|_v = \max_{1\le j\le d+1} \sum_{1\le i\le d+1} |m_{i,j}|.$$
Since these norms are the operator norms associated with the $\ell^\infty$ and the $\ell^1$ norms on ${\mathbb R}^{d+1}$ we have $\|M_1M_2\|_h \le \|M_1\|_h\|M_2\|_h$ and the same for $\|\ ||_v$.

So, we can choose the prime number~$p$ so that

$$p>2\,\max\bigl\{\max_{0\le i< \nu} \|A_i\|_h^n,\max_{0\le i< \nu} \|A_i\|_v^n \bigr\}.$$
\bigskip


\section{Message authentication and integrity}

Now Alice wants to make sure that the message supposedly coming from Bob actually comes from him. They may use the following procedure. Both Alice and Bob will use the second protocol. They share $n$ and~$p$ and the dimension~$k+1$ of the matrices. Alice has already chosen a projective IFS with the separation condition whose keys are
$$
(n,p,B)\text{\quad and\quad} (n,p,A,U,{\mathcal S}).
$$
Bob has already prepared the cryptogram~$C$ to be transmitted.
In his turn he sets up an IFS with matrices of the same dimension of Alice's. His keys are
$$
(n',p,B')\text{\quad and\quad} (n',p,A',U',{\mathcal S}').
$$

Alice choose two words $w'$ and $w''$ on the alphabet $\{0,1,\dots,\nu'-1\}$ both of length $m$ with $m\le n'/2$. We suppose that some bits in $w'w''$  code the number~$m$. She computes $\Phi_{B',w'}$ and $\Phi_{B',w''}$ and sends  $\Phi_{B',w'w''} = \Phi_{B',w'}\Phi_{B',w''}$ to Bob who then is able to extract $w'w''$ and check if the length~$m$ agrees with the information contained in~$w'w''$. If it is so, it is unlikely that there was tampering. Then he sends
$C'=\Phi_{B',w'}C\Phi_{B',w''}$ to Alice who then recovers~$C$: $C =\Phi_{B',w'}^{-1}C'\Phi_{B',w''}^{-1}$.
If the decoding of~$C$ is successful, it is likely that $C'$ has not been tampered. Since Bob was the only one who can easily get $w'w''$ this is him that emitted the cryptogram~$C$.

\end{document}